\documentclass[%
 aip,
rsi,%
 amsmath,amssymb,
 reprint,floatfix%
]{revtex4-1}

\usepackage{graphicx}
\usepackage{amssymb}
\usepackage{amsmath}

\usepackage{array}
\usepackage{float}
\newcolumntype{P}[1]{>{\centering\arraybackslash}p{#1}}
\newcolumntype{M}[1]{>{\centering\arraybackslash}m{#1}}
\usepackage{multirow}

\usepackage{dcolumn}
\usepackage{bm}
\usepackage[normalem]{ulem}
\usepackage{amsmath}
\usepackage{color}

\begin{document}

\preprint{AIP/123-QED}

\title{Interaction of particles with a cavitation bubble near a solid wall}

\author{Leonel A. Teran}
\affiliation{Research Group of Fatigue and Surfaces, Mechanical Engineering School, Universidad del Valle, Cali, Colombia} 
\author{Sara A. Rodriguez} \email{sara.rodriguez@correounivalle.edu.co}
\affiliation{Research Group of Fatigue and Surfaces, Mechanical Engineering School, Universidad del Valle, Cali, Colombia} 
\author{Santiago La\'in} 
\affiliation{Energetics and Mechanics Department, Universidad Aut\'onoma de Occidente, Cali, Colombia} 
\author{Sunghwan Jung} \email{sunnyjsh@cornell.edu}
\affiliation{Department of Biomedical Engineering and Mechanics, Virginia Tech, Blacksburg, Virginia 24061, USA} 
\affiliation{Department of Biological \& Environmental Engineering, Cornell University, Ithaca, New York 14853, USA} 

\date{\today}

\begin{abstract}

Hard particle erosion and cavitation damage are two main wear problems that can affect the internal components of hydraulic machinery such as hydraulic turbines or pumps. If both problems synergistically act together, the damage can be more severe and result in high maintenance costs. In this work, a study of the interaction of hard particles and cavitation bubbles is developed to understand their interactive behavior. Experimental tests and numerical simulations using computational fluid dynamics (CFD) were performed. Experimentally, a cavitation bubble was generated with an electric spark near a solid surface, and its interaction with hard particles of different sizes and materials was observed using a high-speed camera. A simplified analytical approach was developed to model the behavior of the particles near the bubble interface during its collapse. Computationally, we simulated an air bubble that grew and collapsed near a solid wall while interacting with one particle near the bubble interface. Several simulations with different conditions were made and validated with the experimental data. The experimental data obtained from particles above the bubble were consistent with the numerical results and analytical study. The particle size, density and position of the particle with respect to the bubble interface strongly affected the maximum velocity of the particles.
\end{abstract}

\pacs{Valid PACS appear here}

\maketitle


\section{Introduction}
\label{S:Intr}

\begin{figure*}[th]
\centering
\includegraphics[width=1\textwidth]{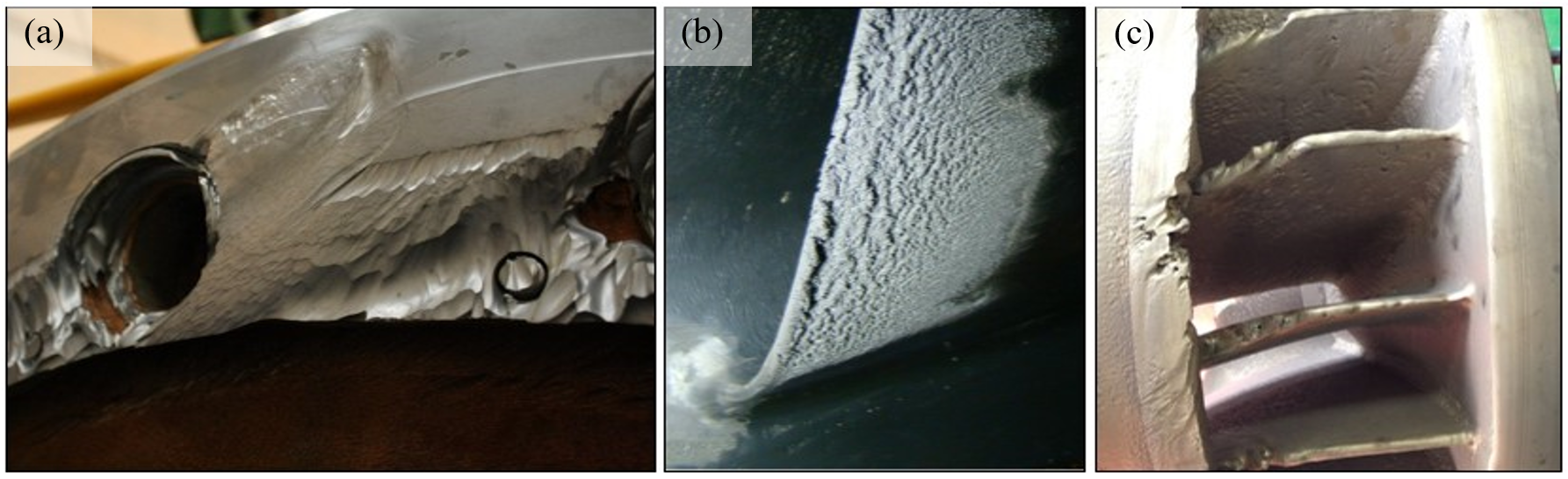}
\caption{\centering \label{fig:SurfaceDamage} Wear damage in the components of three Francis turbines: (a) Hard-particle erosion in one of the covers; (b) Cavitation damage at the trailing edge of one of the runner's blades; (c) Synergistic damage of the hard particles and cavitation at the leading edge of the runner's blades}
\end{figure*}

A general problem of hydraulic machinery such as water pumps and turbines is the wear of their internal components. Some of those components that are in contact with a liquid are susceptible to several problems such as solid particle erosion or cavitation damage \cite{dorji2014hydro,avellan2004introduction,kumar2010study,neopane2010sediment}. For particle erosion, hard particles travel at high speeds in a liquid and hit a solid surface, which causes deformation or loss of material in many cases. The level of damage depends on various factors such as velocity, attack angle, density, hardness, size, concentration, fracture toughness and shape factor of the particles in the liquid \cite{hutchings1992tribology,zambrano2018mild}. The cavitation damage is a product of the collapse of many cavitation bubbles near a solid surface. Cavitation occurs when the liquid pressure due to hydrodynamic effects drops below the vapor pressure of the liquid at a given temperature. Consequently, cavitation bubbles appear and travel with the liquid; when they reach a high-pressure region, they collapse and generate smaller bubbles and pressure waves. These collapses near a solid surface generate micro jets and high-pressure waves, which directly hit the surface and induce deformation. Such deformation continues to increase due to the successive collapses of bubbles, and the surface material is eventually removed due to fatigue \cite{franc2006fundamentals, kim2014advanced}. In some cases, both phenomena act together in a synergistic way, causing more severe damage than that caused by each phenomenon alone \cite{amarendra2012synergy,li2006cavitation,tang2012research,tao2011mechanism,zhang2011experimental}. However, in experiments performed using a vibratory apparatus, particles under a critical size inhibit the cavitation damage instead of increasing the damage \cite{lian2018effect}. Therefore, the study of the interaction of particles and cavitation bubbles is a key aspect to avoid the damage caused by the synergistic effect of cavitation and particles. Fig. \ref{fig:SurfaceDamage} shows the damage of some Francis turbine components due to the three aforementioned phenomena.

Several authors showed that the interaction of particles and cavitation bubbles is caused by the bubble dynamics during their growth and collapse. Soh and Willis \cite{soh2003flow} developed an experiment to observe the movement of several particles due to the collapse of a bubble. In the experiment, they fixed several particles with cords above a bubble, which was generated by an electric spark near a solid wall, and placed others on the solid wall. They found that the suspended particles were not apparently affected because of the restriction of the cords; however, the particles on the surface were significantly moved from their original position. In 2015, Poulain et al. \cite{poulain2015particle} analyzed the dynamics of a spherical particle affected by a cavitation bubble of a similar size, which was generated by an electric spark far from any wall. In their observations, they defined three phases: first, the bubble grows and pushes the particle away; then, the bubble collapses, and the particle is attracted to it; finally, a significant time after the collapse, the particle continues moving toward the center of the bubble due to the rebounding of the cavitation bubble at an earlier time. Using an analytic model based in the asymmetric dynamics of the bubble, the authors found that the normalized velocity of the particle showed an inverse-fourth-power-law relationship with the normalized distance between the bubble and the particle.

The behavior of a cavitation bubble without the interaction of particles has also been studied using CFD numerical simulations. Osterman et al. \cite{osterman2009numerical} studied the collapse of one single bubble in an ultrasonic field near a wall surface. In their analysis, they used a finite-volume 2D axisymmetric model and the Volume-Of-Fluid approach to evaluate the effect of the initial bubble distance from the wall. They validated their simulation with the experimental results of Philip and Lauterborn \cite{philipp1998cavitation} by obtaining consistent shapes of the bubble during the collapse. Their study established that the results were highly sensible to the grid density, and the velocities of the micro jets developed during the collapse were approximately 100 m/s for a maximum bubble radius of 1.45 mm and a separation of 1.74 mm from the wall. Johnsen and Colonius \cite{johnsen2009numerical} simulated the collapse of a gas bubble induced by shock wave in a free field near a solid surface. To validate the simulations, they compared their results with the available theory and experiments an showed consistency with the bubble dynamics data and propagation of the shock emitted upon the collapse. The induced shock collapse generated notably high velocities in the re-entrant jet, which created a water hammer shock that produced notably high pressures on the wall and could represent potential damage to the neighboring surface. Jayaprakash et al. \cite{jayaprakash2012numerical} performed experimental tests and numerical simulations of the interaction of a vertical wall and a bubble. In this case, they simulated the growth and collapse of the bubble using a high initial pressure inside a very small bubble. They validated their results with the data obtained from an experimental bubble generated by an electrical spark. Their studies also showed a notable good correlation between numerical simulations and experimental observations. They concluded that the jet characteristics strongly depended on the standoff distance from the wall.  

If there are particles near a cavitation bubble collapsing near a solid wall, the generated re-entrant jet can trap them in its velocity field; in some conditions, the particles can be accelerated towards the surface and cause damage on it. To evaluate this effect, Li \cite{li2006cavitation} proposed a microscopic model that assumed a low concentration of particles suspended in the fluid and a small particle size compared with the cavitation bubbles; thus, the solid particles did not affect the bubbles or flow characteristics. According to this author, the particle is trapped by the jet of the collapsing bubble and accelerated to a high velocity toward the solid surface. In a later investigation, Dunstan and Li \cite{dunstan2010cavitation} numerically studied the dynamics of only one particle near a cavitating bubble during its collapse. As a result, they verified that the damage potential on a surface was increased because the particle acquired high kinetic energy due to its interaction with the collapsing bubble.

In this work, experimental tests and CFD numerical simulations are used to analyze the behavior of particles immersed in a fluid field generated by the collapse of a cavitation bubble near a solid wall. The experimental tests help to understand the bubble dynamics effect on small particles and also to validate the numerical simulations. These results can help to understand the synergistic damage caused by particles and cavitation, which frequently appear in hydraulic machinery such as water pumps and water turbines.

\section{Experimental tests}
\label{SS:Exp tests}
\begin{figure*}[th]
\centering
\includegraphics[width=1\textwidth]{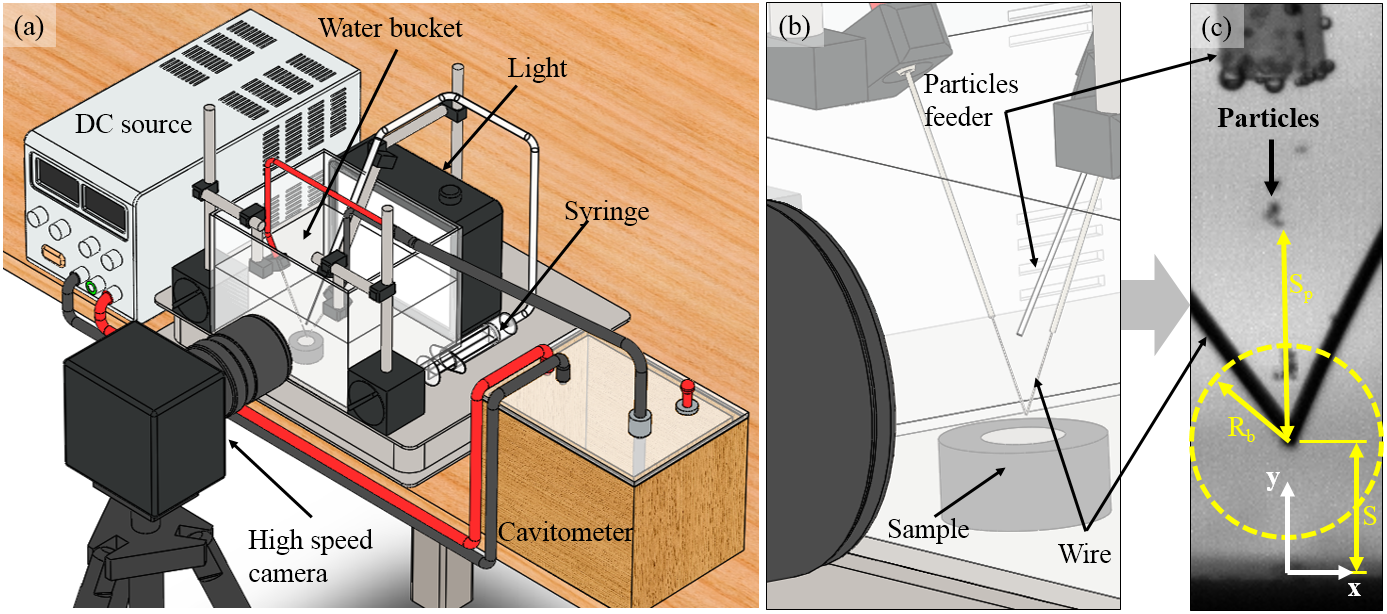}
\caption{\centering\label{fig:Exp_set} Experimental set up: (a) General view; (b) Magnification of the zone where the bubble is generated; (c) Schematic representation of the coordinated system and variables in the experiment}
\end{figure*}

Fig. \ref{fig:Exp_set}(a,b) shows the experimental setup to study the interaction of particles and a cavitation bubble. A schematic of the generated bubble is shown in Fig. \ref{fig:Exp_set}(c) with a reference coordinate centered on the solid surface. A cavitation bubble was created using an electric spark, which was generated when an electric current passes through the tips of two electrodes in contact. A tin cooper wire of an approximate diameter 0.12 mm was used to generate the spark at the desired location. The wire was obtained from a stranded hook-up wire manufactured by Consolidated Electronic Wire \& Cable (Part \# 815-5). The electric energy was obtained from a direct-current (DC) source, which enabled one to change the voltage. Then, the energy was stored in a device composed of 4 capacitors with an equivalent capacitance of 23.5 mF, which could be charged to 50 V. To produce the spark, the stored energy was released to the electrodes and a short circuit was produced in the contact zone and generated a spark that originated the nucleation of a bubble; different voltages resulted in different bubble maximum sizes. A stream of particles was flowing above the nucleation position. The particles were contained in a syringe connected to a tube with a hollow needle (particle feeder) at the end; the tip of the needle and the wires were in the same plane. The particles were transported to the desired position by gravity; once the particles reach that position, the bubble was generated to enable their interaction. The bubble was generated near a solid surface (sample) to observe the behavior of the particles that interacted with a cavitation bubble that collapsed near a solid surface. This process was recorded using a Photron FASTCAM Mini APX RS high-speed camera at 50,000 frames per second, which enabled us to track the particles to obtain important variables such as the particle velocity and acceleration.

As listed in Table \ref{table:ExperimentSet}, four parameters were evaluated in our experiment: distance from a wall, particle size, particle material, and particle-bubble separation. Each configuration was repeated 3 times, and some were also used to validate the CFD simulations. Table \ref{table:ExperimentSet} shows the low and high levels of the evaluated factors in a full factorial experiment. The distance from a wall ($S$ in Fig. \ref{fig:Exp_set}) is the distance between the solid wall and the nucleation center. We tested two particle sizes and two types of particle material. Here, the two materials have different densities: 2130 $\pm$ 210 kg/m$^3$ for sand and 3910 $\pm$ 350 kg/m$^3$ for alumina \cite{zambrano2018mild}. The particle separation parameter ($S_p$) is the distance between a particle above the bubble and the bubble nucleation center, which was measured at its maximum radius. During this experiment, the voltage was maintained constant at 38.2 V, which created a maximum radius of 2.5 $\pm$ 0.06 mm.  
\begin{table}[h]
\centering
\resizebox{0.4\textwidth}{!}{%
\begin{tabular}{ M{2.2cm} M{2cm} M{2cm} }
\hline
& \multicolumn{2}{M{3.6cm}}{\textbf{Factor level}}\\
\textbf{Factors} & Low & high \\
\hline
Distance from a wall ($S$) & 1.5 mm & 2.5 mm \\
\\
Particle size & 53-63 $\mu$m  & 75-106 $\mu$m \\
\\
Particle material & Sand & Alumina \\
\\
Particle-bubble separation ($S_p$) & 2.52-2.58 mm & 2.7-2.8 mm \\
\hline
\end{tabular}}
\caption{\label{table:ExperimentSet}Perfomed experiments}
\end{table}

Other experiments were performed to analyze the effect of the maximum bubble size on the behavior of the particles. Four bubble sizes of 1.5 $\pm$ 0.04 mm, 2.5 $\pm$ 0.06 mm, 3 $\pm$ 0.08 mm, and 4 $\pm$ 0.05 mm were generated using charge voltages of 31.5 V, 38.2 V, 43 V and 47.5 V, respectively. In this case, the distance from the wall was fixed at 1.5 mm, the particle separation from the interface was 0.2-0.3 mm, where the $S_p$ varied between 1.7 and 4.3 depending on the maximum bubble size, the particle size was 75-106 $\mu$m, and the particle material was sand. 

\section{Numerical model}
\label{SS:Num model}
A simplified geometry of an axisymmetric fluid domain was built to perform a simplified numerical simulation of the interaction of a solid particle and an air bubble subject to pressure changes. In this case, the air is modeled as an ideal gas, which is different from the real case, where the gas is water vapor, and there is mass transfer in the evolution of the bubble. In the simplified case, the evolution of the air bubble is due to pressure changes.

Several experimental investigations generated their bubbles using a spark \cite{soh2003flow,poulain2015particle}, a focused laser beam \cite{philipp1998cavitation,supponen2015detailed} or an ultrasonic field \cite{lu2013dynamics,suslick2011extreme}. However, the described simplified simulations have been validated with this type of experimental test \cite{johnsen2009numerical,osterman2009numerical}. In this simulation, the phenomenon is assumed to be axisymmetric because experiments with a single cavitation bubble \cite{philipp1998cavitation,zhang2013experiments,jayaprakash2012numerical} show that during a collapse near a surface, a micro jet is developed through the center of the bubble towards the surface and forms an axis of symmetry. Thus, a two-dimensional simulation that spends less computational resource can be used.

The commercial software ANSYS Fluent solver was used to solve Navier-Stokes equations in a transient simulation. This software enables us to use the Volume-Of-Fluid (VOF) model to capture the behavior of the interface between air and liquid water. Moreover, it is possible to create user-defined functions to define variable boundary conditions or add compressibility effects to the liquid and gas phases.   

The governing equations in this simulation are the mass, momentum and energy conservation equations.

The mass conservation equation can be written as:
\begin{equation}
\label{eq:m_c}
\frac{\partial \rho}{\partial t} + \nabla \cdot (\rho \vec{v}) = 0
\end{equation}
where $\rho$ is the fluid density, and $\vec{v}$ is the velocity vector. With the axisymmetric simplification, the mass conservation equation is:
\begin{equation}
\label{eq:am_c}
\frac{\partial \rho}{\partial t} + \frac{\partial}{\partial y}(\rho v_y) + \frac{\partial}{\partial r}(\rho v_r) + \frac{\rho v_r}{r}=0
\end{equation}
where $y$ is the axial coordinate, $r$ is the radial coordinate, $v_y$ is the axial velocity, and $v_r$ is the radial velocity.
The momentum conservation equation is:
\begin{equation}
\label{eq:mom_c} 
\frac{\partial}{\partial t}(\rho \vec{v}) + \nabla \cdot (\rho \vec{v}\vec{v})= -\nabla p + \nabla \cdot (\underline{\underline{\tau}}) + \rho \vec{g} + \vec{F}
\end{equation}
where $p$ is the static pressure; $\rho \vec{g}$ and $\vec{F}$ are the gravitational and external body forces, respectively. The shear stress tensor $\underline{\underline{\tau}}$ is defined by:
\begin{equation}
\label{eq:ss_e}
\underline{\underline{\tau}}=\mu \Big[\Big(\nabla \vec{v}+\nabla \vec{v}^T\Big)-\frac{2}{3}\nabla \cdot \vec{v}I\Big]
\end{equation} 
where $\mu$ is the dynamic viscosity, $I$ is the unit tensor, and the second term of the right-hand side is the effect of volume dilation.

In the axisymmetric case, the axial and radial momentum conservation equations are defined as:
\begin{eqnarray}
\label{eq:ax_momx}
&&\frac{\partial}{\partial t}(\rho v_y) + \frac{1}{r}\frac{\partial}{\partial y}(r\rho v_y v_y) + \frac{1}{r}\frac{\partial}{\partial r}(r\rho v_r v_y) \nonumber \\
& = &-\frac{\partial p}{\partial y} + \frac{1}{r}\frac{\partial}{\partial y}\Big[r\mu \Big(2\frac{\partial v_y}{\partial y}-\frac{2}{3}(\nabla \cdot \vec{v})\Big)\Big] \nonumber \\
&&+\frac{1}{r} \frac{\partial}{\partial r} \Big[r\mu \Big(2\frac{\partial v_y}{\partial r} + \frac{\partial v_r}{\partial y} \Big)\Big] + F_y
\end{eqnarray}

and 
\begin{eqnarray}
\label{eq:ax_momr}
&&\frac{\partial}{\partial t}(\rho v_r) + \frac{1}{r}\frac{\partial}{\partial y}(r\rho v_y v_r) + \frac{1}{r}\frac{\partial}{\partial r}(r\rho v_r v_r) \nonumber \\ 
&= &-\frac{\partial p}{\partial r} + \frac{1}{r}\frac{\partial}{\partial r}\Big[r\mu \Big(2\frac{\partial v_r}{\partial r}-\frac{2}{3}(\nabla \cdot \vec{v})\Big)\Big] \nonumber \\
&&+\frac{1}{r} \frac{\partial}{\partial y} \Big[r\mu \Big(2\frac{\partial v_r}{\partial y} + \frac{\partial v_y}{\partial r} \Big)\Big] \nonumber \\
&&-2\mu \frac{v_r}{r^2} + \frac{2}{3}\frac{\mu}{r}(\nabla \cdot \vec{v})+ F_r
\end{eqnarray}

where
\begin{equation}
\label{eq:div_v}
\nabla \cdot \vec{v} = \frac{\partial v_y}{\partial y} + \frac{\partial v_r}{\partial r} + \frac{v_r}{r}
\end{equation}

The energy conservation equation in its general form can be written as:
\begin{eqnarray}
\label{eq:energy_consv}
&&\frac{\partial}{\partial t} (\rho E) + \nabla \cdot (\vec{v}(\rho E + p)) \nonumber \\
&=&\nabla \cdot \bigg(k\nabla T - \sum _j h_jJ_j + \Big(\underline{\underline{\tau}} \cdot \vec{v} \Big) \bigg) + S_h
\end{eqnarray}
where
\begin{equation}
\label{eq:energy}
E=h-\frac{p}{\rho}+\frac{v^2}{2}
\end{equation}
The first three terms on the right side of  Eq. \eqref{eq:energy_consv} represent the energy transfer by conduction, species diffusion and viscous dissipation, where $k$, $T$, $h$ and $J$ are the thermal conductivity, temperature, sensible enthalpy and diffusion flux of species, respectively. The last term ($S_h$) is a volumetric heat source. In the VOF model, there is no species diffusion, there is no condensation or evaporation, and the heat source is zero.   

The VOF two-phase model describes the behavior of a primary phase in a secondary phase assuming that the phases do not mix with each other. In this study this model is used to observe the behavior of an air bubble (primary phase) in liquid water (secondary phase) \cite{osterman2009numerical}. The tracking of the interface surface between the two phases is obtained by solving the continuity equation for the volume fraction ($\alpha$) of the secondary phase:
\begin{equation}
\label{eq:vof}
\frac{\partial}{\partial t}(\alpha_2\rho_2)+\nabla \cdot(\alpha_2\rho_2\vec{v})=0
\end{equation}
Because of the limitations of the VOF model, the right-hand side of Eq. \eqref{eq:vof} is zero since there is no mass transfer of the source terms of evaporation or condensation. The volume fraction of the primary phase is calculated considering that the sum of the two volume fractions is one. Given the known value of the volume fraction of one phase in a computational cell, the fields for all variables and properties are shared by the two phases and represent a volume-average value at each location. 

The two phases in the VOF model are considered compressible. The primary phase is air, and its properties are defined using the ideal gas law, whereas the density of secondary phase (water) depends on the pressure according to the following expression:
\begin{equation}
\label{eq:water_density}
\rho = \frac{\rho_0}{1-\frac{\Delta p}{K}}
\end{equation}
where $K$ is the water bulk modulus (2.2 GPa), $\rho_0$ is the reference density (1000 kg/m$^3$), and $\Delta p$ is the pressure difference regarding the reference pressure (1 atm) \cite{osterman2009numerical}.

The simulated domain was a rectangle of 200 mm $\times$ 200 mm, which represents a 400-mm-diameter cylinder in the axisymmetric case. The selected size helps to avoid boundary effects in the small region where the phenomenon occurs ( $S/R_b$ ratio of 0.37-1) \cite{johnsen2009numerical}, additionally, in test simulations of domain sizes 50 mm $\times$ 50 mm, 200 mm $\times$ 200 mm and 500 mm $\times$ 500 mm, the results obtained with the last two sizes had a difference in maximum velocity of the micro jet less than 3$\%$. Therefore, to reduce the computational cost, the size of 200 mm $\times$ 200 mm was selected. The domain was discretized using a hexahedral structured mesh, which was refined in the region where the bubble grew and collapsed. A small air bubble of radius 60 $\mu$m was placed on the symmetry axis at the beginning of the simulation near a solid wall at several positions according to the numerical experiment described in table \ref{table:Numericalset}. During the growth of the bubble, a maximum size is achieve, which depends on the initial pressure in the fluid domain to be simulated. 

\begin{figure}[h]
\centering
\includegraphics[clip,width=0.4\textwidth]{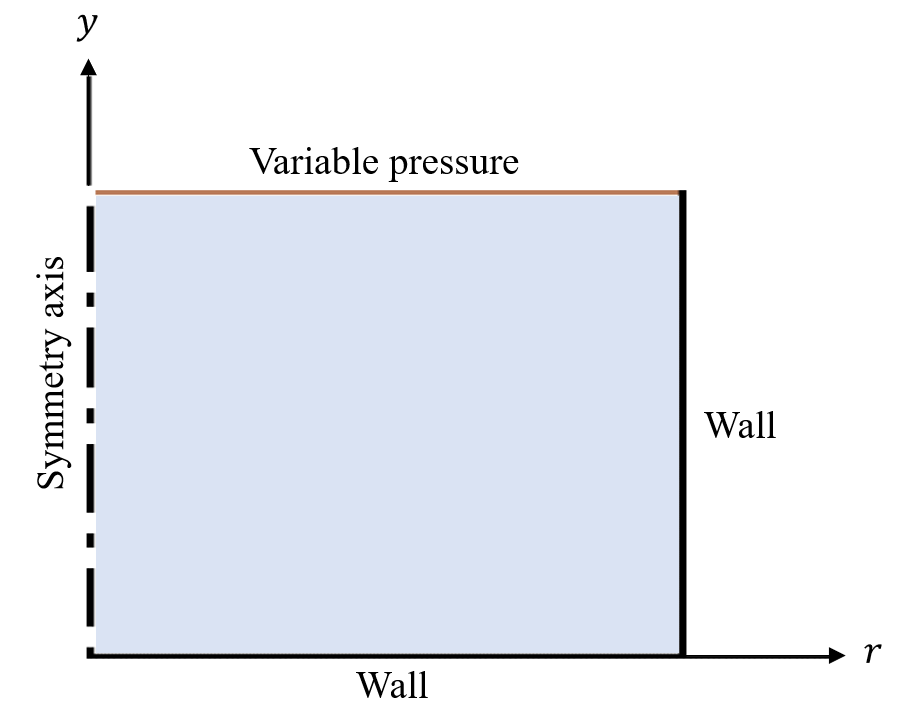}
\caption{\centering \label{fig:BoundaryC} Boundary conditions for the bubble collapse simulation}
\end{figure}

The boundary conditions are shown in Fig. \ref{fig:BoundaryC}. The pressure boundary condition at the top edge of the domain is a step function that varies with time. In this boundary, a initial pressure of 110 kPa immediately decreases to 1 kPa and is maintained during $4 \times 10 ^{-4}$ s; then, the pressure changes to 100 kPa and remains constant until the end of the simulation to enable the bubble collapse. The particle was modeled as a wall boundary with a circular shape and located in the symmetry axis at different positions from the bubble (see table \ref{table:Numericalset}); this particle could move along the symmetry axis. To model the interaction between the particle and the fluid, the Six DOF solver of the ANSYS Fluent software coupled with a dynamic mesh was used, which enabled us to calculate the forces and moments of an object of six degrees of freedom immersed in a fluid.

The pressure and velocity were couple using a coupled scheme that included the volume fractions of air and water. In the spatial discretization, the pressure was interpolated using a PRESTO! scheme \cite{fluent2015theory}, whereas the density, momentum and energy were interpolated using a second-order scheme, and a compressive scheme was used in the case of the volume fraction. Finally, the temporal discretization was set to be first-order implicit using a time step size of $10^{-8}$ s. 
Table \ref{table:Numericalset} shows the performed numerical simulations to validate the results and evaluate several parameters in the interaction of the particle with the cavitation bubble.

Before performing the validation with the experimental data, the mesh size of the proposed simulation in section \ref{SS:Num model} was evaluated in a mesh independence study. Therefore, the maximum cell size to obtain reliable results in the zone where the bubble grows and collapses is a square of side 10 $\mu$m.
   
\begin{table}[h]
\centering
\resizebox{0.5\textwidth}{!}{%
\begin{tabular}{ M{1.6cm} M{1.6cm} M{2.5cm} M{1.6cm} M{1.6cm} }
\hline
\textbf{Particle size} & \textbf{Particle material} & \textbf{$S_p$} & \textbf{$S$} & \textbf{$R_b$}\\
\hline
60 $\mu$m & Sand & 2.52-2.58 mm & 1.5 mm & 2.5 mm \\
60 $\mu$m & Sand & 2.52-2.58 mm & 2.5 mm & 2.5 mm \\
90 $\mu$m & Sand & 2.52-2.58 mm & 2.5 mm & 2.5 mm \\
90 $\mu$m & Alumina & 2.52-2.58 mm & 2.5 mm & 2.5 mm \\
\hline
\end{tabular}}
\caption{\label{table:Numericalset}Performed numerical simulations}
\end{table}

\section{Results and discussion}
\label{S:Results and discussion}
\subsection{Experimental test results}
\label{SS:Exp res}
\begin{figure}[h]
\centering
\includegraphics[width=0.5\textwidth]{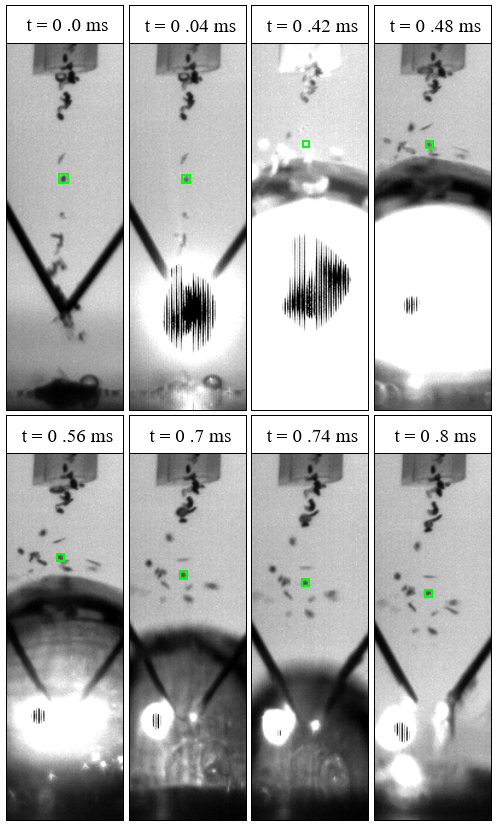}
\caption{\centering \label{fig:BubbEv} Image sequence of the bubble evolution due to the spark for a test, where the particle size is 75-106 $\mu$m, the particle material is sand, the distance from the wall is 1.5 mm, and the maximum size of the bubble is 2.53 mm}
\end{figure}

\begin{figure}[h]
\centering
\includegraphics[width=0.45\textwidth]{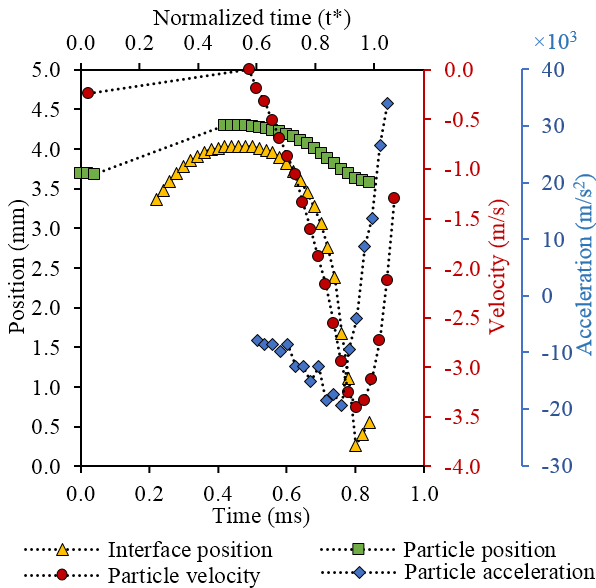}
\caption{\centering \label{fig:Part-int_Vpos} Evolution  of a bubble interface and a particle vertical position, velocity and acceleration for a test, where the particle size is 75-106 $\mu$m, the particle material is sand, the distance from the wall is 1.5 mm, and the maximum size of the bubble is 2.53 mm}
\end{figure}

Fig. \ref{fig:BubbEv} shows a sequence of the evolution of the bubble and particles with one of the tracked particles highlighted. Fig. \ref{fig:Part-int_Vpos} shows the evolution of the vertical position of the highlighted particle and bubble interface, and the particle velocity and acceleration during the growth and collapse of the bubble for testing, where the particle size is 75-106 $\mu$m, the particle material is sand, the distance from the wall is 1.5 mm, and the maximum size of the bubble is 2.53 mm. All analyzed particles were tracked using the software Tracker \cite{trackerSoftware}. The bubble dynamics enables particle movement during the growth (time 0-0.42 ms in Fig. \ref{fig:BubbEv}) and collapse (time 0.48-0.8 ms in Fig. \ref{fig:BubbEv}) of the bubble. It was not possible to observe the particle movement during a long part of the bubble growth due to the bright spark in every test; however, the full collapse process, where the maximum velocities of the micro jet are developed, was well observed. The normalized time $t^*$ in Fig. \ref{fig:Part-int_Vpos} was calculated using the required time of bubble growth and collapse as the reference, which was defined as the instant immediately before the spark appeared ($t_i$) until the moment when the bubble interface reached the initial position of the bubble's center ($t_{gc}$). The expression is:

\begin{equation}
\label{eq:Norm_time}
t^*= \frac{t-t_i}{t_{gc}-t_i}
\end{equation}

\begin{figure*}
\centering
\begin{tabular}{c c}
\includegraphics[width=0.5\textwidth]{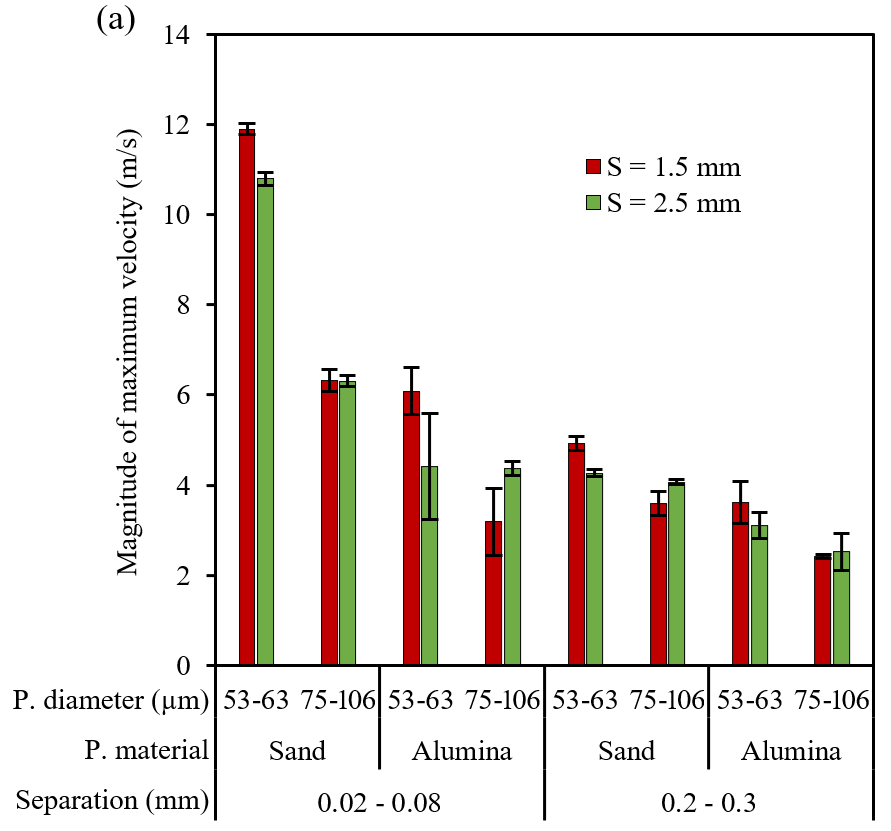} & \includegraphics[width=0.5\textwidth]{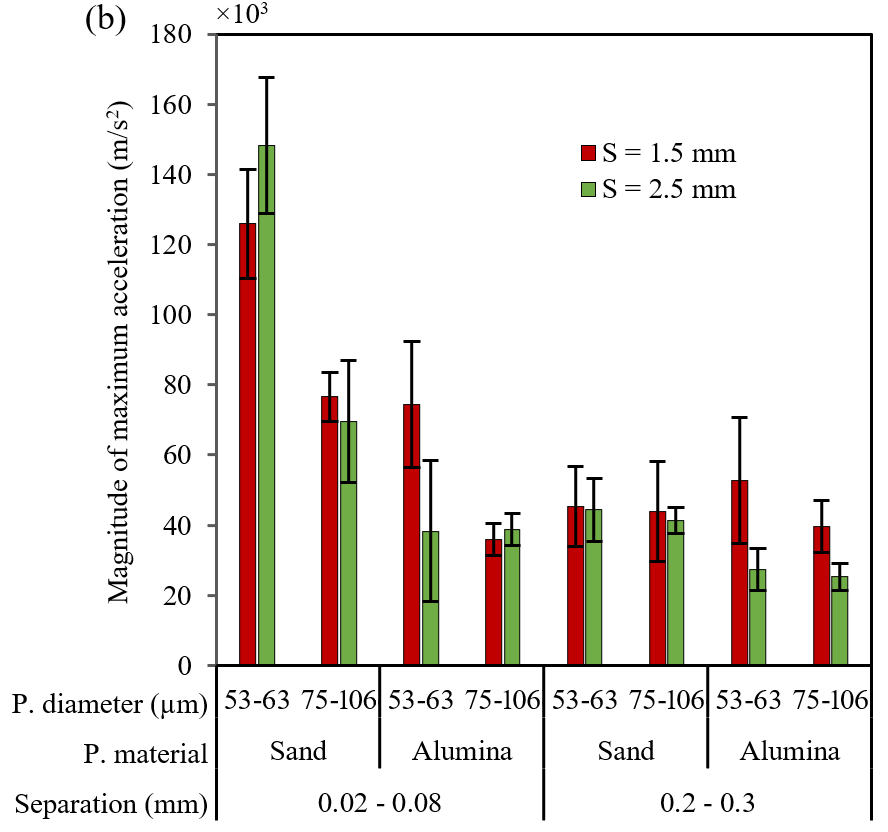}
\end{tabular}
\caption{\centering \label{fig:ExpResults} Experimental results of the particle behavior: a) Magnitude of the maximum velocity; b) Magnitude of the maximum acceleration }
\end{figure*}

Fig. \ref{fig:ExpResults} shows the experimental test results for all evaluated cases. The left and right graphs show the magnitude of the maximum velocity and  maximum acceleration of particles above the bubble in the vertical direction. Neither velocity nor acceleration significantly varied with the distance of the wall in most cases. In fact, an ANOVA study of the performed factorial experiment reveals that the most significant parameters from less to greater relevance in the particle velocity are the particle size, material, and position with respect to the bubble interface. Table \ref{table:ANOVA_Res} shows the F and P values for the main effects and their combined effects on the particle maximum velocity and acceleration; the P values below 0.05 are significant. Thus, both particle size and material significantly affect the particle velocity because of their effect on the particle mass since larger and denser alumina particles (whose density is 3910 kg/m$^3$) are more difficult to move than small and lighter sand particles (whose density is 2150 kg/m$^3$). Moreover, particles near the bubble interface are highly affected by the velocity field generated from the bubble growth and collapse, whereas a smaller effect is observed on the particles far from the bubble interface.

\begin{figure}[h]
\centering
\includegraphics[width=0.46\textwidth]{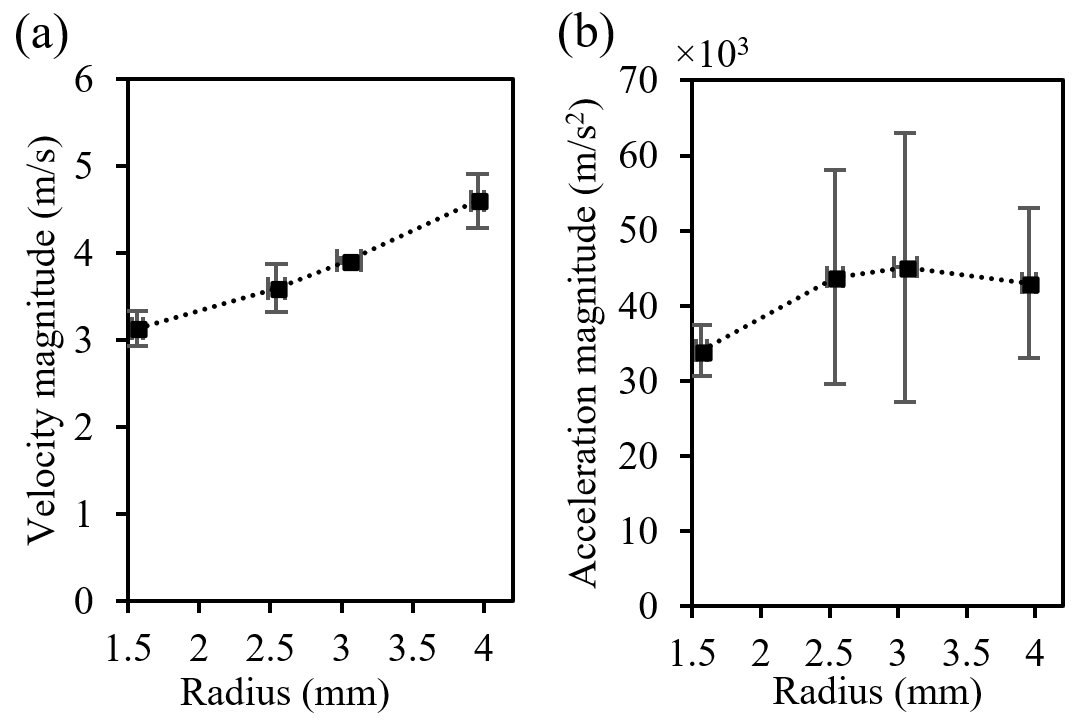}
\caption{\centering \label{fig:bubSize} Experimental evaluation of bubble size on the particle behavior. a) Magnitude of the maximum velocity; b) Magnitude of maximum acceleration. In this case, the distance from the wall was 1.5 mm, the particle separation from the interface was 0.2-0.3 mm, the particle size was 75-106 $\mu$m, and the particle material was sand}
\end{figure}

Fig. \ref{fig:bubSize} presents the behavior of the particles with the variation in the maximum bubble size for sand particles of size 75-106 $\mu$m; the initial position of the bubble from the wall was 1.5 mm, and the separation of the bubble interface was 0.2-0.3 mm. An increase in the bubble radius escalates the maximum velocity of the particles above the bubble, whereas the maximum acceleration is not influenced significantly.

\begin{table}[h]
\centering
\resizebox{0.48\textwidth}{!}{%
\begin{tabular}{ m{5.2cm} M{1cm} M{1cm} M{1cm} M{1cm}}
\hline
& \multicolumn{2}{M{2.1cm}}{\textbf{Maximum velocity}} & \multicolumn{2}{M{2.1cm}}{\textbf{Maximum acceleration}}\\ 
\textbf{Source} & F & P & F & P \\
\hline
$S$ & 6.34 & 0.017 & 5.5 & 0.025 \\
Particle size & 350.18 & 0 & 51.06 & 0\\
Particle material & 663.12 & 0 & 102.06 & 0\\
$S_p$ & 814.01 & 0 & 122.3 & 0\\
$S$*Particle size & 42.35 & 0 & 0.56 & 0.46\\
$S$*Particle material & 0.2 & 0.661 &	10.61 & 0.003\\
$S$*$S_p$ &	1.35 & 0.253 & 0.94 & 0.34\\
Particle size*Particle material & 62.18 & 0 & 9.28 & 0.005\\
Particle size*$S_p$ & 123.64 & 0 & 31.53 & 0\\
Particle material*$S_p$ & 192.02 & 0 & 60.77 & 0\\
$S$*Particle size*Particle material & 2.08 & 0.159 & 9.62 & 0.004\\
$S$*Particle size*$S_p$ & 6.36 & 0.017 & 0 & 0.995\\
$S$*Particle material*$S_p$ & 0.95 & 0.336 & 0.23 & 0.636\\
Particle size*Particle material*$S_p$ & 72.05 &	0 & 14.97 & 0.001\\
$S$*Particle size*Particle material*$S_p$ & 6.88 & 0.013 & 4.62 & 0.039\\
\hline
\end{tabular}}
\caption{\centering \label{table:ANOVA_Res}F and P values from the ANOVA study for the maximum velocity and maximum acceleration}
\end{table}

\subsection{Validation of CFD simulations}
\label{SS:Exp Validation}
A simplified analytical approach, which considers the spherical collapse of a bubble and away from any solid surface, was developed to understand the behavior of the particles above the bubble during the collapse phase. In this analysis, equations \eqref{eq:velocCyl}-\eqref{eq:Particle_Velocity_1} were adapted from Poulain et al.\cite{poulain2015particle}. Considering that the flow created by the evolution of the bubble is incompressible \cite{plesset1977bubble} and using spherical coordinates centered at the nucleation site, radial velocity $u$ can be written as:
\begin{equation}
\label{eq:velocCyl}
u(r,t)=\Big(\frac{R_b(t)}{r}\Big)^2\dot{R}_b(t)
\end{equation}
where $R_b$ is the bubble radius. Considering negligible gravitational force, an analysis of the forces that act on the particle because of the flow leads to the following expression:
\begin{equation}
\label{eq:partForc}
m_p\ddot{r}_p=F_{drag}(t)
\end{equation}
where $m_p$ is the particle mass, $r_p$ is the radial position of the particle, $\ddot{r}_p$ is the particle acceleration, and $F_{drag}(t)$ is the drag force, which is defined as:
\begin{equation}
\label{eq:dragForce}
F_{drag}(t)=\text{sgn}(u)\frac{1}{2}C_D\rho \pi R_p^2 [u(r_p(t),t)]^2
\end{equation}
$R_p$ is the particle radius, $\rho$ is the density of water, and $C_D$ is the drag coefficient, which is assumed constant at a value of 0.47 according to the experimental data reported by \citet{NASA2010SphereDrag}.
The particle velocity can be evaluated by integrating Eq. \eqref{eq:partForc} over the bubble collapse phase. As a simplification, a negligible displacement is assumed in the drag expression because the particle does not experience a significant movement, as shown in Fig. \ref{fig:Part-int_Vpos}. Thus, $r_p=r_{p,i}$, where $r_{p,i}$ is the initial position of the particle immediately before the bubble begins to collapse. Therefore, the particle velocity can be written as:
\begin{equation}
\label{eq:Particle_Velocity_1}
\dot{r}_p(t)=\text{sgn}(u)\frac{\rho C_D\pi R_p^2}{2m_p r_{p,i}^4}\int_{t_i}^{t} [R_b(t)]^4 [\dot{R}_b(t)]^2 dt
\end{equation}
where $t_i$ is the initial time; $\dot{R}_b$ is the velocity of the bubble interface, which is a function of time and can be evaluated using Rayleigh-Plesset equation \cite{franc2006fundamentals}
\begin{eqnarray}
\label{eq:Rayleigh-Plesset}
&&\rho\bigg[R_b\ddot{R}_b + \frac{3}{2}\dot{R}_b^2\bigg]\nonumber\\
&=& p_v - p_{\infty}(t)+p_{g0}\Big(\frac{R_0}{R_b}\Big)^{3\gamma}-\frac{2S}{R_b}-4\mu \frac{\dot{R}_b}{R_b}
\end{eqnarray}
Here, $p_v$ is the vapor pressure of water at the operating temperature; $p_{\infty}$ is the pressure in the bulk of the surrounding liquid, which is a function of time; $p_{g0}$ is the initial partial pressure of the gas inside the bubble; $R_0$ is the initial radius of the bubble; $\gamma$ is the ratio of the heat gas capacities $c_{pg}$ and $c_{vg}$; $S$ is the surface tension.

In the case of the bubble collapse, for simplicity, the effects of the viscosity, non-condensable gas and surface tension are assumed to be negligible. Therefore, Eq. \eqref{eq:Rayleigh-Plesset} can be integrated to yield
\begin{equation}
\label{eq:RayleighSimplified}
\rho \dot{R}_b^2R_b^3 = -\frac{2}{3}(p_{\infty}(t)-p_v)(R_b^3-R_0^3)
\end{equation}
During the collapse of the bubble, $\dot{R}$ is negative, so
\begin{equation}
\label{eq:Interfacevelocity}
\dot{R}_b=-\sqrt{\frac{2}{3}\frac{p_{\infty}-p_v}{\rho}\bigg[\frac{R_0^3}{R_b^3}-1\bigg]}
\end{equation}
Combining Eqs. \eqref{eq:Interfacevelocity} and \eqref{eq:Particle_Velocity_1} leads to
\begin{equation}
\label{eq:particle_velocity_2}
\dot{r}_p(t)=\frac{C_D \pi R_p^2(p_{\infty}-p_v)}{3 m_p r_{p,i}^4}\int_{t_i}^t(R_0^3R_b-R_b^4)dt
\end{equation}
In this analysis, Eq. \eqref{eq:Interfacevelocity} was solved using numerical integration with the trapezoidal rule to find the collapse time and evaluate the evolution of the bubble radius with time. Then, this solution was used to solve Eq. \eqref{eq:particle_velocity_2} using numerical integration to evaluate the particle velocity. After several tests of convergence, the integration interval was divided in 2000 sub-intervals to achieve reliable results.

The previous analysis helps to understand the behavior of the particle because of the bubble dynamic effect during the collapse phase; however, there is no bubble after the collapse, so the particle begins to decelerate because of the drag caused by the fluid, which is assumed to be static, until the particle stops as observed in the experiments. To evaluate the particle velocity after this phase of the movement, the momentum equation of the particle was used to obtain
\begin{equation}
\label{eq:final_drag momentum}
\ddot{r}_p=-\frac{\rho C_D \pi R_p^2}{2m_p}\dot{r}_p^2
\end{equation}

The integration of the equation \eqref{eq:final_drag momentum} results in
\begin{equation}
\label{eq:velocity after collapse}
\dot{r}_p(t)=\frac{2m_p\dot{r}_{p,i}}{2m_p+\rho C_D \pi R_p^2 \dot{r}_{p,i} t}
\end{equation}
which enables us to evaluate the particle velocity after the bubble collapse to compare with experimental data.

\begin{figure}[h]
\centering
\includegraphics[width=0.46\textwidth]{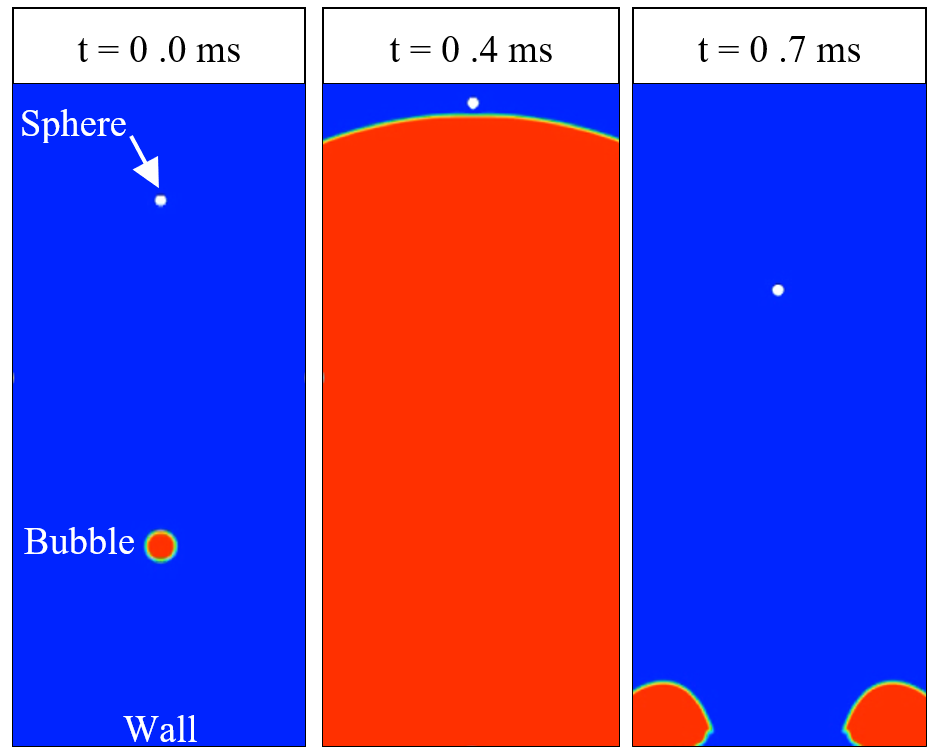}
\caption{\centering \label{fig:CFDsequ} Image sequence of an air bubble evolution obtained with a CFD simulation}
\end{figure}

\begin{figure*}
\centering
\includegraphics[width=1\textwidth]{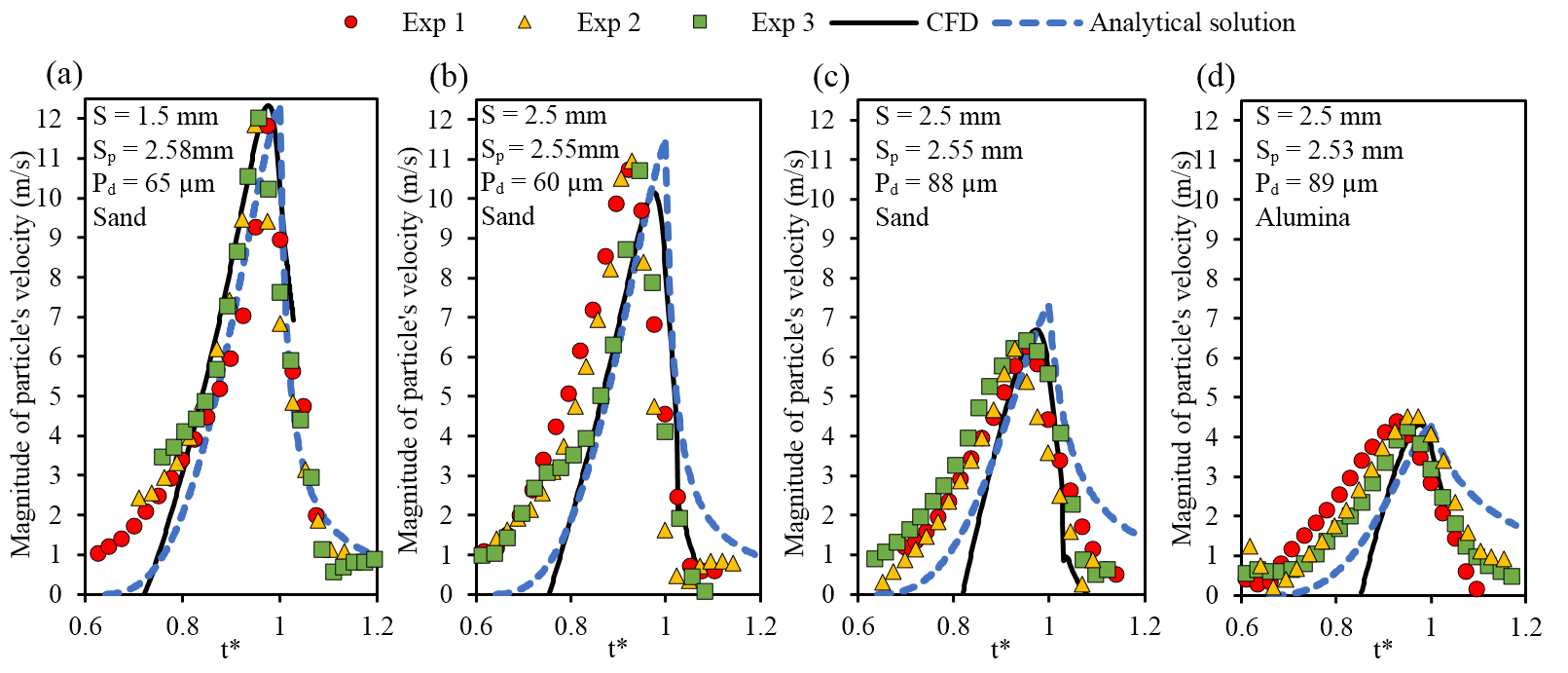}
\caption{\centering \label{fig:CFD_comp} Experimental validation of CFD simulation for a maximum bubble size ($R_b$) of 2.5 mm: a) Distance to wall: 1.5 mm; particle diameter: 65 $\mu$m; particle material: sand; b) Distance to wall: 2.5 mm; particle diameter: 60 $\mu$m; particle material: sand; c) Distance to wall: 2.5 mm; particle diameter: 88 $\mu$m; particle material: sand; d) Distance to wall: 2.5 mm; particle diameter: 89 $\mu$m; particle material: alumina. The particle diameter in the experimental tests was estimated as an equivalent diameter}
\end{figure*}

Fig. \ref{fig:CFDsequ} presents a sequence of CFD result images of the volume fraction of an air bubble (red color) and its interaction with a particle (white) during the growth and collapse in water (blue). Fig. \ref{fig:CFD_comp} shows the magnitude of the particle velocity in several graphs to compare the experimental data with numerical CFD and the analytical solutions obtained with Eqs. \eqref{eq:particle_velocity_2} and \eqref{eq:velocity after collapse} in several conditions during the bubble collapse. In the Fig. \ref{fig:CFD_comp}  we used the normalized time of Eq. \eqref{eq:Norm_time} to better compare the results. 

The CFD simulations predict notably well the behavior of the particle affected by the bubble dynamics, particularly after the particle reaches its maximum velocity. The under-prediction at the beginning of the collapse can be a result of the differences between the CFD and experimental initial position of the particle, which was difficult to identify due to the bright spark at the beginning of the experiment. The analytical analysis provides a good prediction of the particle's behavior above the bubble and under the conditions of a spherical collapse despite all assumed simplifications. This analysis helps to identify two phases for the behavior of a particle immersed in the field near the bubble during its collapse. In the first phase, which begins at the maximum radius of the bubble, the particle velocity increases and reaches its maximum when the bubble radius is near zero. In the second phase, the fluid was assumed static as observed in the high-speed images, and the velocity of the particle decreases, which indicates that the fluid decelerates the particle.

\section{Conclusion}
Experimental and numerical approaches were used to study the interaction of particles with a cavitation bubble. This analysis helps to understand the effect of the bubble dynamics on particles of different sizes and densities. When the particles are above the bubble, their density and size have a strong effect, whereas the bubble separation from the surface was not significant for the results of maximum velocity. Additionally, the bubble size has a small effect on the velocity of the particle located above the bubble. The experimental results and a simple analytical study validate the numerical simulations. Further experimental and numerical studies will be performed to investigate the potential damage of a particle accelerated by the collapse of a cavitation bubble on a metallic surface. 

\section*{Acknowledgements}
The authors acknowledge Colciencias, Universidad del Valle, and Virginia Tech for the support during the development of this project. Also, this research is partially supported by the National Science Foundation (CBET-1604424).

\section*{Bibliography}
\bibliographystyle{model1-num-names}
\bibliography{sample.bib}







\end{document}